\documentclass[aps,physrev,reprint,superscriptaddress]{revtex4-2}

\usepackage{graphicx}
\usepackage{todonotes}
\usepackage{amsmath}
\usepackage{amsfonts}
\usepackage{mathrsfs}
\usepackage{xcolor}
\usepackage[colorlinks=true, allcolors=blue]{hyperref}
\newcommand{\bea}{\begin{eqnarray}}
\newcommand{\eea}{\end{eqnarray}}
\newcommand{\nn}{\nonumber \\}

\def\mean#1{\langle#1\rangle}

\begin{document}

\preprint{APS/123-QED}

\title{Multi-Modal Spectroscopy Theory for Ultrafast Control of Rabi Oscillations}% Force line breaks with \\
\author{Jia-Wang Yu}
%\email{jw.yu@zju.edu.cn}
\affiliation{State Key Laboratory of Silicon and Advanced Semiconductor Materials \textnormal{\&} College of Information Science and Electronic Engineering, Zhejiang University, Hangzhou 310027, China}
\affiliation{ZJU-Hangzhou Global Scientific and Technological Innovation Center, Zhejiang University, Hangzhou 311200, China}
\author{Hong-Bin Wang}
\thanks{Jia-Wang Yu and Hong-Bin Wang contributed equally to this work.}
\affiliation{State Key Laboratory of Silicon and Advanced Semiconductor Materials \textnormal{\&} College of Information Science and Electronic Engineering, Zhejiang University, Hangzhou 310027, China}
\affiliation{ZJU-Hangzhou Global Scientific and Technological Innovation Center, Zhejiang University, Hangzhou 311200, China}
\author{Xiao-Qing Zhou}
\affiliation{Department of Physics, School of Science, Westlake University, Hangzhou 310030, China}
\affiliation{Institute of Natural Sciences, Westlake Institute for Advanced Study, Hangzhou 310024, China}
\author{Min Tang}
\affiliation{School of Optoelectronics Engineering and Instrumentation Science, Dalian University of Technology, Dalian, 116024, China}
\author{Zhi-Bo Ni}
\affiliation{State Key Laboratory of Silicon and Advanced Semiconductor Materials \textnormal{\&} College of Information Science and Electronic Engineering, Zhejiang University, Hangzhou 310027, China}
\affiliation{ZJU-Hangzhou Global Scientific and Technological Innovation Center, Zhejiang University, Hangzhou 311200, China}
\author{Xiao-Tian Cheng}
\affiliation{State Key Laboratory of Silicon and Advanced Semiconductor Materials \textnormal{\&} College of Information Science and Electronic Engineering, Zhejiang University, Hangzhou 310027, China}
\affiliation{ZJU-Hangzhou Global Scientific and Technological Innovation Center, Zhejiang University, Hangzhou 311200, China}
\author{Yi Zhao}
\affiliation{State Key Laboratory of Silicon and Advanced Semiconductor Materials \textnormal{\&} College of Information Science and Electronic Engineering, Zhejiang University, Hangzhou 310027, China}
\affiliation{ZJU-Hangzhou Global Scientific and Technological Innovation Center, Zhejiang University, Hangzhou 311200, China}
\author{Shu-Ning Ding}
\affiliation{State Key Laboratory of Silicon and Advanced Semiconductor Materials \textnormal{\&} College of Information Science and Electronic Engineering, Zhejiang University, Hangzhou 310027, China}
\affiliation{ZJU-Hangzhou Global Scientific and Technological Innovation Center, Zhejiang University, Hangzhou 311200, China}
\author{Jun-Yong Yan}
\affiliation{State Key Laboratory of Silicon and Advanced Semiconductor Materials \textnormal{\&} College of Information Science and Electronic Engineering, Zhejiang University, Hangzhou 310027, China}
\author{Hui-Hui Zhu}
\affiliation{State Key Laboratory of Silicon and Advanced Semiconductor Materials \textnormal{\&} College of Information Science and Electronic Engineering, Zhejiang University, Hangzhou 310027, China}
\affiliation{ZJU-Hangzhou Global Scientific and Technological Innovation Center, Zhejiang University, Hangzhou 311200, China}
\affiliation{College of Integrated Circuits, Zhejiang University, Hangzhou 311200, China}
\author{Chen-Hui Li}
\affiliation{State Key Laboratory of Silicon and Advanced Semiconductor Materials \textnormal{\&} College of Information Science and Electronic Engineering, Zhejiang University, Hangzhou 310027, China}
\affiliation{ZJU-Hangzhou Global Scientific and Technological Innovation Center, Zhejiang University, Hangzhou 311200, China}
\author{Feng Liu}
\affiliation{State Key Laboratory of Silicon and Advanced Semiconductor Materials \textnormal{\&} College of Information Science and Electronic Engineering, Zhejiang University, Hangzhou 310027, China}
\author{Chao-Yuan Jin}
\email{jincy@zju.edu.cn}
\affiliation{State Key Laboratory of Silicon and Advanced Semiconductor Materials \textnormal{\&} College of Information Science and Electronic Engineering, Zhejiang University, Hangzhou 310027, China}
\affiliation{ZJU-Hangzhou Global Scientific and Technological Innovation Center, Zhejiang University, Hangzhou 311200, China}
\affiliation{College of Integrated Circuits, Zhejiang University, Hangzhou 311200, China}

%%%%%%%%%%%%%%%%%%%%%%%%%%%%%%%%%%%%%%%%%%%%%%%


%affiliation{
%\affiliation{
 %Third institution, the second for Charlie Author
%\affiliation{%


\date{\today}% It is always \today, today,

\begin{abstract}
% full control
We propose a three-cavity scheme to realize full control of the emitter-cavity coupling strength in cavity quantum electrodynamics (cQED). The involvement of coupled oscillators gives rise to transient dynamics comprising multiple spectral components, which significantly increases the numerical cost to resolve the fluorescence spectrum in the time domain. A generalized sensor method is hence developed to simplify the calculation process for the characterization of nonstationary quantum dynamics. Multi-modal spectroscopy reveals the emergence, splitting, and disappearance of supermodes in real time. Based on the depletion of the zero-energy supermode, ultrafast switching of Rabi oscillations is demonstrated for time-domain multi-modal spectroscopy. These results exhibit a consistent picture from the spectral control of multiple oscillators to the quantum observation in ultrafast dynamics, which establishes the sensor method as a powerful theoretical tool for the ultrafast spectroscopy of cQED systems. 

\end{abstract}

\maketitle

\section{Introduction}
% Coherent control of light-matter interaction lies at the heart of the modern development of quantum optics and quantum information technologies, whereas the deterministic preparation of quantum states critically relies on real-time manipulation of interaction strength \cite{shields2007semiconductor, kimble2008quantum, raimond2001manipulating, cirac1997quantum}. To this end, we have proposed non-local control schemes based on multi-cavity architectures to modulate the emitter-cavity interaction without directly perturbing the quantum emitter \cite{johne2015control, jin2014ultrafast, pellegrino2018deterministic}. The cavity systems can be naturally utilized for quantum logic gates, where multiple optical modes participate coherently in the dynamics, and the coupling strengths can largely exceed the dissipation rates. In this regime, the energy exchange between the quantum emitter and the cavity optical field takes place on timescales comparable to or shorter than the decoherence time. Understanding and characterizing the dynamical response of such systems therefore requires access to time-resolved spectroscopic information. However, the coexistence of ultrafast temporal evolution and multiple spectral components inherent to multi-mode cavity systems demands high resolution in both time and frequency. This requirement, together with the time-energy uncertainty relations in the measurement process, makes the direct calculation and interpretation of transient spectrum particularly challenging \cite{yamaguchi2022time}.

Coherent control of light-matter interaction lies at the heart of the modern development of quantum optics and quantum information technologies, whereas the deterministic preparation of quantum states critically relies on real-time manipulation of interaction strength \cite{shields2007semiconductor, kimble2008quantum, raimond2001manipulating, cirac1997quantum}. In cavity quantum electrodynamics (cQED) systems, the ability to deterministically tailor the coupling strength between the atomic two-level system (TLS) and the cavity optical field is particularly important for the applications of solid-state single-photon sources and quantum logic gates. Several technologies have been proposed and developed to dynamically tune the frequency of cavity modes in and out of the atomic transition frequency \cite{faraon2010fast, bose2014all, albert2013microcavity, shambat2011ultrafast, hogele2004voltage, faraon2007local, midolo2013controlling, thyrrestrup2013non}. Such schemes are typically dependent on the frequency shift of cavity resonance via thermal control \cite{faraon2009local}, electrical control \cite{faraon2010fast, petruzzella2015fully}, or optomechanical modulation \cite{tian2022all, midolo2018nano}, which inevitably perturbs the quantum emitter itself. One method avoiding local perturbation in single-photon sources is introduced by Johne et al. \cite{johne2015control, jin2014ultrafast, pellegrino2018deterministic}. In their proposal, three coupled cavities were utilized to enable non-local modulation of quantum vacuum field experienced by a single-photon emitter without directly perturbing its transition frequency. 

Apart from single-photon devices, TLS-three-cavity systems can be naturally utilized for quantum logic gates, whereas multiple optical modes participate coherently in quantum dynamics and the coupling strengths largely exceed the dissipation rates. In this regime, the energy exchange between the quantum emitter and the cavity optical field takes place on timescales comparable to or shorter than the decoherence time. Understanding and characterizing the dynamical response of such systems therefore requires access to time-resolved spectroscopy. However, the coexistence of ultrafast temporal evolution and multiple spectral components inherent to multi-mode cavity systems demands high resolution in both time and frequency domains. This requirement, together with the time-energy uncertainty relations in the measurement process, makes the direct calculation and interpretation of transient spectrum particularly challenging \cite{yamaguchi2022time}.

Although the transient spectrum of a simple TLS-cavity system can be computed using the quantum regression theorem (QRT) \cite{yamaguchi2022time}, its evaluation in general time-dependent systems requires the calculation of two-time correlation functions and the repeated numerical evaluation of a double time integral over a range of frequencies. Achieving high temporal and spectral resolutions typically requires fine discretization in the time domain, as well as dense sampling over the frequency domain, which substantially increases the computational cost.

These challenges are particularly pronounced in multi-mode and nonstationary systems relevant to ultrafast light-matter control, where resolving rapid dynamical features requires fine discretization in both time and frequency domains \cite{yu2026enhancing, wigger2021resonance, karpinski2021control}. Numerous experimental demonstrations of ultrafast light-matter control in cQED systems therefore call for an efficient and easy-to-use numerical framework capable of computing transient spectrum in nonstationary regimes \cite{peinke2021tailoring, sattler2020probing, jin2014ultrafast, fuhrmann2011dynamic}. 
In this context, the transient spectrum serves as a key diagnostic observable that reveals how externally applied control protocols manifest in the system's dynamical response.

In this work, we present a formulation of the sensor method \cite{del2012theory, holdaway2018perturbation, ruiz2014spontaneous} for computing TRPS \cite{yamaguchi2022time,eberly1977time,jin2014ultrafast} in nonstationary cQED systems governed by time-dependent Hamiltonians. The method provides an efficient and physically transparent approach to multi-modal transient spectroscopy and remains applicable beyond the steady-state regime. As a representative example, we apply the method to a three-cavity system, which naturally exhibits multiple coupled modes and enables ultrafast modulation of light-matter interaction via cavity detuning. In this system, the control mechanism is governed by the redistribution of excitation among different supermodes. The transient physical spectrum provides a direct way to resolve this redistribution in both time and frequency, thereby revealing the underlying dynamical processes. Applying the method to the three-cavity system, we show how different spectral resolutions capture the time-dependent redistribution of vacuum-field modes. Our results establish the sensor method as an effective numerical tool for transient spectroscopy in systems with complex modal structures and provide a theoretical foundation for its application in future ultrafast quantum optical experiments.

\section{Ultrafast switching of Rabi oscillation in multi-cavities}

Before introducing the time-resolved spectroscopy framework, we first examine the dynamical response of the three-cavity system under ultrafast switching. This allows us to identify the underlying control mechanism, which will later be characterized using the transient physical spectrum.

\subsection{Multi-modal Hamiltonian}

\begin{figure}[!h]
    \centering
    \includegraphics[width=0.45\textwidth]{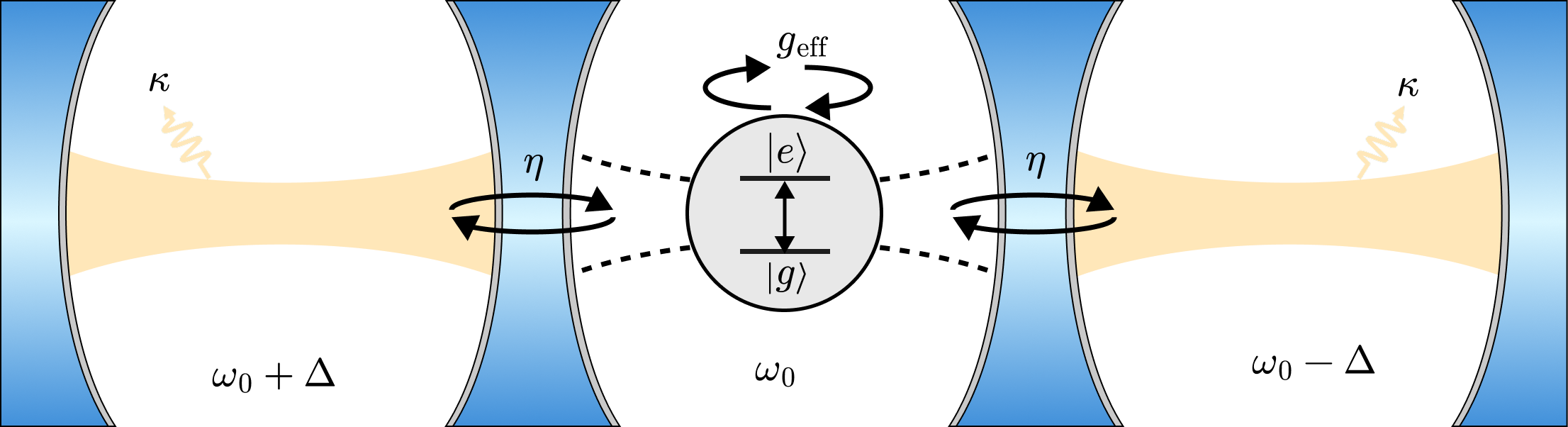}
    \caption{
    Schematic illustration of the TLS-three-cavity system. A two-level system (TLS) is embedded in the middle cavity and couples to its local photonic mode with strength $g$. The two lateral cavities are coupled to the middle cavity with strength $\eta$, and their resonance frequencies are tuned to $\omega_0+\Delta$ and $\omega_0-\Delta$, respectively. The supermode of the three-cavity system at $\omega_0$ is mainly distributed over the lateral cavities and nearly vanishes in the middle cavity when $\Delta = 0$ (yellow shaded region). In contrast, when $|\Delta| \gg \eta$, the $\omega_0$ mode becomes localized in the middle cavity (black dashed outline). This detuning-controlled redistribution of the optical density of states (DOS) in the middle cavity enables modulation of the effective coupling strength $g_{\mathrm{eff}}$ between the TLS and supermode.
    }
    \label{fig:threecavity}
\end{figure}

The subject of study consists of a TLS coupled to three cavities, as illustrated in Fig.~\ref{fig:threecavity} \cite{johne2015control,konoike2019chip}. The TLS is embedded in the middle cavity and interacts with its local photonic mode with strength $g$, while the two lateral cavities are coupled to the middle cavity with strength $\eta$. All coupling strengths are assumed to be much larger than dissipation rates. The system Hamiltonian is written as
\begin{equation}
H_S = H_0 + H_I ,
\end{equation}
where the free Hamiltonian is
\begin{equation}
\label{eq:H0}
H_0 = \omega_e\, \sigma^\dagger \sigma
+ \omega_l a^\dagger a
+ \omega_m b^\dagger b
+ \omega_r c^\dagger c ,
\end{equation}
and the interaction Hamiltonian under the rotating-wave approximation (RWA) is
\begin{equation}
\label{eq:HI}
H_I = g(\sigma^\dagger b + \sigma b^\dagger)
+ \eta (b^\dagger a + b^\dagger c + a^\dagger b + c^\dagger b).
\end{equation}

Here, $a$, $b$, and $c$ denote the annihilation operators of the left, middle, and right cavities with resonance frequencies $\omega_l$, $\omega_m$, and $\omega_r$, respectively, while $\sigma$ is the lowering operator of the TLS with transition frequency $\omega_e$. The first term in $H_I$ describes the coupling between the TLS and the optical field in the middle cavity, while the second term represents the coupling of the optical field between the middle cavity and two lateral cavities. 

We consider the system dynamics restricted to the single-excitation subspace, which is sufficient to capture the coherent interaction between the TLS and the cavity modes in the present regime. We impose the resonant condition $\omega_e = \omega_m = \omega_0$ for the TLS and middle cavity, and apply symmetric detuning to the lateral cavities, $\omega_l = \omega_0 + \Delta$ and $\omega_r = \omega_0 - \Delta$. The Hamiltonian then takes the form
\begin{equation}
\label{eq:matrixH}
H_S =
\begin{pmatrix}
\omega_0 & 0 & g & 0 \\
0 & \omega_0+\Delta & \eta & 0 \\
g & \eta & \omega_0 & \eta \\
0 & 0 & \eta & \omega_0-\Delta
\end{pmatrix},
\end{equation}
where the lower-right $3\times 3$ block corresponds to the three-cavity subsystem, denoted as $H_{\mathrm{cav}}$.

$H_{\mathrm{cav}}$ can be diagonalized analytically, yielding the eigenfrequencies
\begin{equation}
\label{eq:eigenenergies}
\omega_0,\qquad
\omega_{\pm} = \omega_0 \pm \sqrt{\Delta^2 + 2\eta^2},
\end{equation}
where $\omega_0$ corresponds to the central supermode (zero-energy mode), and $\omega_-$ and $\omega_+$ correspond to the lower and upper supermodes, respectively. The corresponding normalized eigenvectors are
\begin{widetext}
\begin{equation}
\label{eq:3cavEvec}
\vec{v}_0 =
\frac{1}{\sqrt{2+(\Delta/\eta)^2}}
\begin{pmatrix}
-1 \\[2mm]
\Delta/\eta \\[2mm]
1
\end{pmatrix},\quad
\vec{v}_- =
\frac{1}{l_-}
\begin{pmatrix}
\displaystyle\frac{-\Delta\sqrt{\Delta^2+2\eta^2} + \Delta^2 + \eta^2}{\eta^2} \\[2mm]
\displaystyle\frac{-\sqrt{\Delta^2+2\eta^2}+\Delta}{\eta} \\[2mm]
1
\end{pmatrix},\quad
\vec{v}_+ =
\frac{1}{l_+}
\begin{pmatrix}
\displaystyle\frac{\Delta\sqrt{\Delta^2+2\eta^2} + \Delta^2 + \eta^2}{\eta^2} \\[2mm]
\displaystyle\frac{\sqrt{\Delta^2+2\eta^2}+\Delta}{\eta} \\[2mm]
1
\end{pmatrix},
\end{equation}
\end{widetext}
where $l_{\pm}$ are the normalization constants given in Appendix~\ref{app:supermodes}.  
The eigenvector $\vec{v}_0$ corresponds to the zero-energy supermode, whose eigenfrequency remains exactly $\omega_0$ and is therefore independent of $\Delta$.

The Hamiltonian on the basis of the supermodes and the TLS is then given by
\begin{align}
H_{dia}&=&\left(\begin{array}{cccc}
	\omega_0&g v_{-}^{(2)} &  g v_{0}^{(2)}&g v_{+}^{(2)}\nn
	gv_{-}^{(2)}& \omega_{-} &0&0\nn
	gv_{0}^{(2)}&0&\omega_0&0\nn
	gv_{+}^{(2)}&0&0&\omega_{+}
\end{array}\right).
\end{align}

\begin{figure}[!t]
    \centering
    \includegraphics[width=0.45\textwidth]{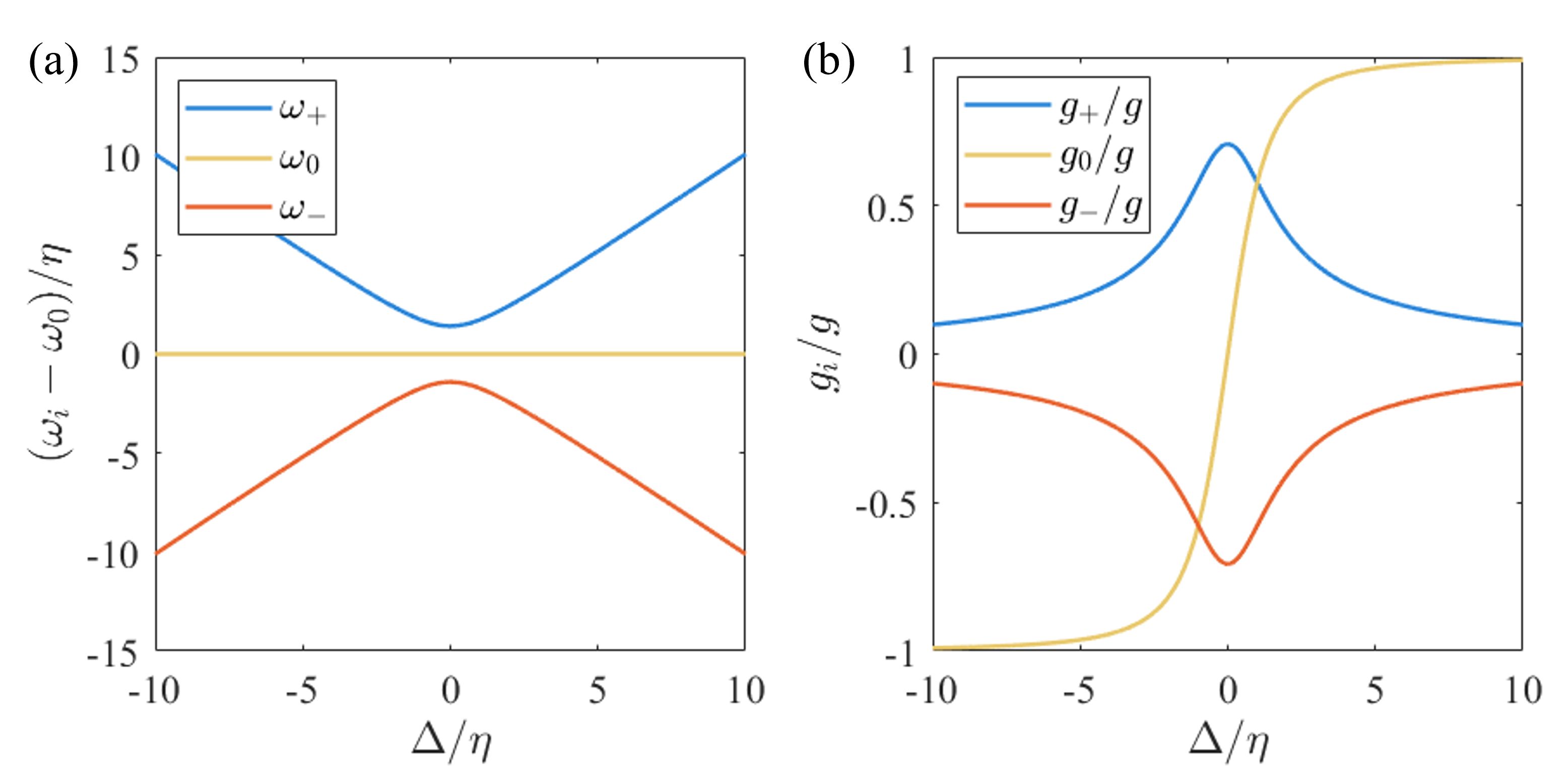}
    \caption{
    (a) Eigenfrequencies of the three supermodes as functions of the normalized detuning $\Delta/\eta$.  
    The central supermode remains pinned at $\omega_0$ and is independent of $\Delta$, whereas the upper and lower supermodes split symmetrically with increasing $|\Delta|$.  
    (b) Effective coupling strengths between the TLS and the three supermodes. In particular, the TLS-central-supermode coupling strength $g_{\mathrm{eff}}$ can be tuned continuously from $-g$ to $g$ by varying $\Delta$, offering a significantly large tuning range.}
    \label{fig:eigenvalues}
\end{figure}

Fig.~\ref{fig:eigenvalues}(a) shows how the frequencies of three supermodes vary with the detuning $\Delta/\eta$.
When $\eta \gg g$, the upper and lower supermodes $\vec{v}_{\pm}$ remain far off-resonant from the TLS, exhibiting a detuning of $\sqrt{\Delta^{2}+2\eta^{2}}$, which increases further as $|\Delta|$ grows. Fig.~\ref{fig:eigenvalues}(b) presents the corresponding effective coupling between the TLS and supermodes. The TLS couples to the upper and lower supermodes only through the reduced strengths $g|v_{\pm}^{(2)}| < g$, while these modes remain strongly detuned in the regime $\eta \gg g$ even at $\Delta = 0$. Under these conditions, their dynamical contribution is negligible under RWA. The TLS-supermode coupling involving $\vec{v}_{\pm}$ is therefore strongly suppressed. When $g$ becomes comparable to $\eta$, however, this separation of scales no longer holds, and the residual coupling to the upper and lower supermodes leads to observable deviations from the suppressed Rabi dynamics (see Appendix~\ref{app:g_eta}). As a result, the Rabi dynamics of the TLS-cavity system is dominated solely by the zero-energy mode. From $H_{dia}$, the TLS couples to the zero-energy mode with an effective strength
\begin{equation}\label{eq:geff}
g_{\mathrm{eff}} = g\, v_0^{(2)}
= g\, \frac{\Delta/\eta}{\sqrt{2+(\Delta/\eta)^2}}.
\end{equation}
The dependence of $g_{\mathrm{eff}}$ on $\Delta$ is shown in Fig.~\ref{fig:eigenvalues}(b), which illustrates that $g_{\mathrm{eff}}$ can be tuned continuously from $-g$ to $g$ \cite{konoike2019chip}. 
This wide tuning range represents a substantial advantage of the three-cavity system compared with the coupled two-cavity non-local control scheme discussed in Ref.~\cite{jin2014ultrafast}.

\subsection{Ultrafast control of coupling strength}
\begin{figure*}[!t]
	\centering
	\includegraphics[width=0.95\linewidth]{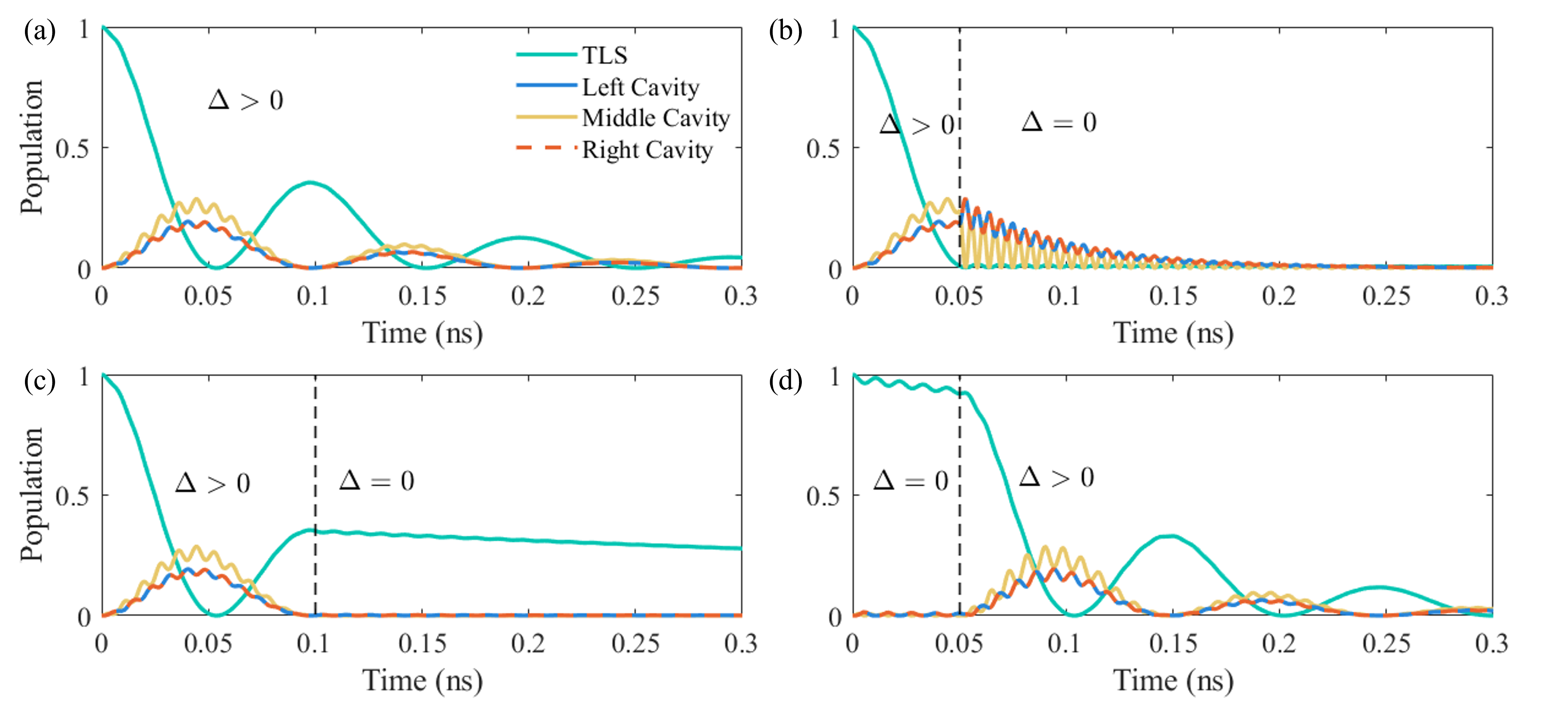}
	\caption{Numerical result for  the Rabi oscillation of  the occupation of the excited state of TLS and population of the three modes. The system is initially at the excited state of the TLS. In (a), the detuning is not changing during the entire process. In (b) and (c) the detuning is turned from $1.2\eta$ to 0 at the vertical dashed line,  while in (d) the detuning is turned from 0 to $1.2\eta$. The other parameters including coupling constants and dissipation rates are given in the text. }
	\label{fig:rabi_abcd}
\end{figure*}

As shown in the previous section, frequency detuning $\Delta$ governs the spatial profile of supermodes and thereby controls the effective coupling strength $g_{\mathrm{eff}}$ between the TLS and the zero-energy supermode. In particular, $g_{\mathrm{eff}}$ vanishes at $\Delta=0$ and approaches $g$ when $|\Delta|\gg \eta$, enabling dynamical on-off control of light-matter interaction simply by modulating $\Delta$. Crucially, varying $\Delta$ also reshapes the spatial distribution of the supermodes. When $\Delta = 0$, the zero-energy mode is supported entirely by the two lateral cavities and sustains a zero field in the middle cavity which de-coupled from the TLS. In contrast, when $|\Delta|\gg \eta$, the field weight of zero-energy mode becomes predominantly concentrated in the middle cavity and the overlap with the TLS maximizes. In time-domain dynamics, a sudden change in $\Delta$ therefore modifies the field composition of supermodes $\vec{v}_0$, $\vec{v}_-$ and $\vec{v}_+$. The decomposition of field weights in terms of instantaneous supermodes changes abruptly. In this picture, the switching operation induces a sudden redistribution of excitation among the supermodes, which leads to qualitatively different transient dynamics.

To investigate the population dynamics under ultrafast control, we simulate the system Hamiltonian $H_S$ using a master equation approach with Lindblad operators. The system is initialized in the state $|e000\rangle$, where the TLS is excited and the cavity modes is unoccupied. During the time evolution, the detuning $\Delta$ is dynamically tuned as $1.2\eta$ or 0 to control the coupling strength. Numerical simulations are performed using the parameter set $\{g,\eta,\gamma,\kappa\} = \{50\text{GHz}, 400\text{GHz}, 1\text{GHz}, 20\text{GHz}\}$, where $\gamma$ and $\kappa$ denote the dissipation rates of the TLS and cavities, respectively. The resulting dynamical evolution in the presence of ultrafast control is shown in Fig.~\ref{fig:rabi_abcd}. It illustrates the time-dependent populations of the excited state of TLS and the cavity modes, clearly revealing the switching of the characteristic Rabi oscillations. 

% The modulation range of the detuning is from $0$ to $4\text{THz}$ immediately by the thermal tuning. Since the thermal tuning rate is fast enough compared to the dissipation rates and coupling  \cite{jin2014ultrafast}, it is implied that the adjustment of the  detuning can be treated as an instantaneous process.

Fig.~\ref{fig:rabi_abcd}(b)-(c) show that switching the detuning at different phases of the Rabi oscillation can either arrest the oscillation or induce pronounced beating in the cavity populations. The qualitative outcome is primarily determined by whether the system is switched into a regime where the TLS effectively couples to a single near-resonant mode or into a regime where multiple supermodes participate in the dynamics. In the present parameter regime $\eta \gg g$, the TLS dynamics is mainly governed by its coupling to the zero-energy mode, characterized by an effective interaction strength $g_{\mathrm{eff}} = g\,v_{0}^{(2)}$. The upper and lower supermodes lie at frequencies $\omega_{\pm}=\omega_0\pm\sqrt{\Delta^2+2\eta^2}$ and therefore become increasingly off-resonant as $|\Delta|$ grows. As a result, their influence on the TLS dynamics is reduced to small corrections, while the dominant energy exchange occurs between the TLS and the zero-energy mode. 
By contrast, when the system is tuned to $\Delta=0$, the mode structure changes so that the TLS no longer couples efficiently to the zero-energy mode and is instead influenced by the coherent superposition of another two supermodes, leading to pronounced multimode interference and beating.

\section{Time-resolved spectroscopy theory} 

The control mechanism in the three-cavity system is governed by the redistribution of excitation among different supermodes. However, the contributions of these supermodes to the dynamics are not directly reflected in population observables alone. The TRPS provides a direct means to access this information. In this section, we first introduce the time-resolved spectroscopy framework and the corresponding formulation of the TRPS, then describe the sensor method for its efficient computation, and finally present the resulting multi-modal spectral dynamics.

TRPS provides access to the real-time formation of spectral features that cannot be inferred from population dynamics or steady-state measurements \cite{eberly1977time, roman2018time, christensen2018driving, wigger2021resonance}. 
This transient spectral evolution underlies phenomena such as the temporary suppression of Rabi \cite{yamaguchi2022time} or Autler-Townes splittings \cite{bai1986transient}  and the appearance of mode-selective oscillations in dynamically modulated emitters \cite{moelbjerg2012resonance}. In practical measurements, the detected signal is always shaped by the finite temporal and spectral response of the apparatus, and the use of frequency filters in particular further constrains the accessible information \cite{castro2025time, kuruma2018time}. These considerations make it essential to employ a framework that explicitly accounts for the measurement process when characterizing ultrafast dynamics.

Fundamentally, any realistic detection event reflects the causality and time-energy uncertainty of quantum processes \cite{yamaguchi2022time}. Recognizing these measurement-induced constraints is therefore crucial when transient emission signals are involved. TRPS naturally incorporates these effects and provides a faithful characterization of how spectral features emerge and evolve with time. Efficient theoretical tools for computing TRPS are thus indispensable for interpreting experiments and optimizing quantum optical systems operating in nonstationary regimes.

\begin{figure}
    \centering
    \includegraphics[width=0.45\textwidth]{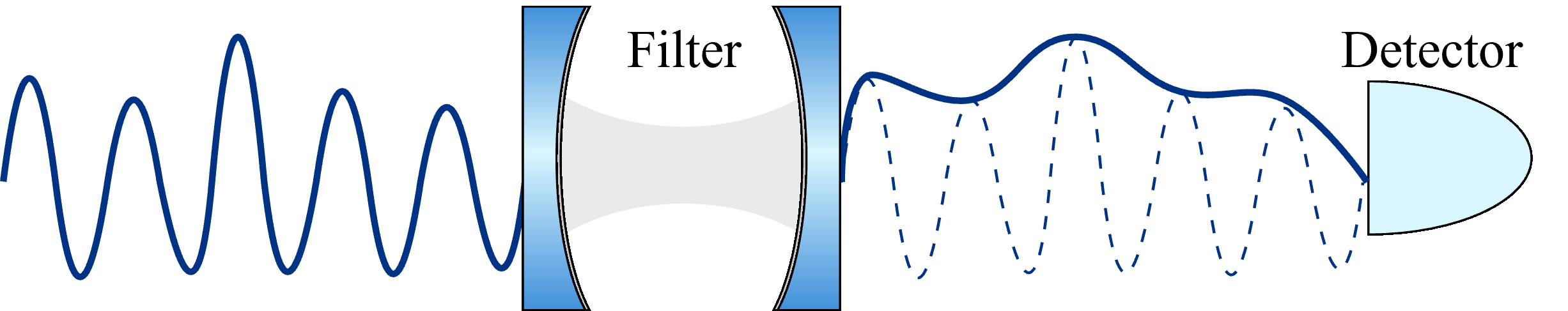}
    \caption{
        Schematic illustration of a time-resolved physical spectrum (TRPS) measurement. 
        The emitted field is passed through a Lorentzian spectral filter, and the time-dependent intensity of the filtered signal is recorded by a sensor, yielding a frequency-resolved temporal profile of the emission.
    }\label{fig:TRPS}
\end{figure}

\subsection{Sensor method}
The physical spectrum is introduced by Eberly and Wodkiewicz~\cite{eberly1977time}, where the finite linewidth of a realistic sensor is explicitly taken into account in the measurement of transient spectrum.
The TRPS of mode $O$ with spectral resolution $\delta_s$ can be written as \cite{eberly1977time,yamaguchi2022time,del2012theory}
\begin{equation}\label{eq:SoPhysicalSpectrum}
\begin{aligned}
S_O(\omega, t, \delta_s)
=&\frac{\delta_s}{\pi} \int^{t}_{-\infty }dt' e^{-\delta_s (t-t')} 
\Re\int_{0}^{\infty}d\tau\, e^{-\frac{\delta_s}{2}\tau} e^{-i\omega\tau}\\
&\times\langle O^\dagger(t')\,O(t'-\tau)\rangle .
\end{aligned}
\end{equation}

Computing the TRPS requires evaluating two-time correlation functions under explicitly time-dependent conditions. The first step is to compute the correlator $\langle O^\dagger(t')O(t'-\tau)\rangle$ over a range of physical times $t'$ and delay times $\tau$, which typically requires propagating the system dynamics for each delay and leads to a numerical cost scaling as $\mathcal{O}(N_t N_\tau)$, where $N_t$ is the number of time steps and $N_\tau$ is the number of delay points.

The TRPS further involves a double time integration over both the physical time $t'$ and the delay time $\tau$, as indicated in Eq.~(\ref{eq:SoPhysicalSpectrum}). In addition to evaluating the correlator on a two-dimensional $(t',\tau)$ grid, the weighted integration over both variables must be carried out for each observation time and each frequency of interest. In particular, for each observation time $t$ and frequency $\omega$, the TRPS is obtained by performing a weighted integration over both the physical time $t'$ and the delay time $\tau$, including a Fourier transform along the $\tau$ direction. This requires $\mathcal{O}(N_t N_\tau)$ operations per frequency point, so that scanning over $N_\omega$ frequency samples leads to an overall computational cost that scales as $\mathcal{O}(N_t N_\tau N_\omega)$, where $N_\omega$ is the number of sampled frequency points.

This computational overhead becomes increasingly demanding in nonstationary or multi-mode systems, where fine resolution in both time and frequency is required, motivating the development of more efficient and physically transparent approaches.

In this work, we present a formulation of the sensor method for explicitly time-dependent situations, building on its established validity for steady-state spectral calculations \cite{del2012theory, holdaway2018perturbation, ruiz2014spontaneous}. In stationary situations, the relevant correlation functions depend only on the time delay $\tau$. In contrast, here we consider nonstationary dynamics with a general (possibly time-dependent) Hamiltonian $H_S(t)$, where the two-time correlation functions take the form $\langle O^\dagger(t')\,O(t'-\tau)\rangle$ and depend explicitly on both the physical time $t'$ and the delay $\tau$. Accordingly, the present formulation does not assume stationary dynamics and can therefore be directly applied to time-dependent systems. Within this formulation, weakly coupled sensors act as frequency-selective probes that track the instantaneous emission, allowing us to reliably reproduce the TRPS of general quantum systems beyond the stationary regime.

A useful physical interpretation of the sensor method is that  weakly coupled auxiliary modes act as minimally invasive, frequency-selective sensors. Each mode acts as a resonant probe. When tuned to a particular frequency component of the spectrum, it allows excitation at that frequency to leak into the probe without appreciably disturbing the system's intrinsic dynamics. The population of each probe thus directly reflects the spectral weight within its frequency band, analogous to an ideal narrowband filter followed by a sensor. By assembling multiple sensors with different center frequencies, one obtains a time-resolved spectral decomposition of the emission that closely mirrors the operational principle of TRPS. A formal justification of this equivalence is provided in Appendix~\ref{app:sensormethodproof}.

Operationally, each sensor is modeled as a weakly coupled auxiliary mode with a center frequency $\omega_k$ and linewidth $\Gamma$, enabling it to function as a narrowband filter that selectively absorbs excitation from the system. The resulting joint Hamiltonian takes the form
\begin{equation}\label{eq:Htot}
H_{\mathrm{tot}}(t) = H_S(t) + \omega_k \zeta_k^\dagger \zeta_k + \epsilon_k \left( \zeta_k O^\dagger + \zeta_k^\dagger O \right), 
\end{equation}
where $H_S(t)$ denotes the system Hamiltonian, and the second term describes the free Hamiltonian of a single sensor mode. The third term represents the weak coupling between the sensor $\zeta_k$ and the system operator $O$, namely $\sigma,a,b,c$ in the three-cavity scheme and their arbitrary linear combinations. Explicitly, the sensor selectively absorbs excitations from the mode $O$ near its center frequency $\omega_k$. When the coupling strength $\epsilon_k$ is chosen sufficiently small, the sensor does not perturb the intrinsic dynamics of the system, yet its instantaneous population faithfully encodes the spectral content within its frequency window.

By solving the dynamics of the joint system governed by $H_{\mathrm{tot}}(t)$, the populations of the sensor modes $n_k=\zeta_k^\dagger \zeta_k$ take the form of the TRPS evaluated at the sensor frequencies $\omega=\omega_k$ with a spectral resolution set by the sensor linewidth $\delta_s=\Gamma$,
\begin{equation}\label{eq:nkSensorPhySpectra}
  \mean{n_k(t)} = \frac{2\epsilon^2\pi}{\Gamma}\,S_O(\omega_k, t, \Gamma).
\end{equation}
In the limit of vanishing sensor linewidth $\Gamma\to 0$, corresponding to an ideal sensor with infinite spectral resolution, Eq.~(\ref{eq:nkSensorPhySpectra}) reduces to the Wiener-Khintchine spectrum, in agreement with the discussion in Ref.~\cite{eberly1977time}. This provides a clear justification for the widely used probability amplitude method, in which the steady-state spectrum is inferred from the amplitudes of perfectly narrowband “sensor modes” such as free-space or waveguide modes in the long-time limit~\cite{qamar2009atom,johne2011single}.

In practice, the TRPS is obtained by scanning the sensor frequency $\omega_k$ over the desired range and performing independent dynamical simulations for each frequency point. For a fixed sensor frequency, the evolution of the joint system governed by $H_{\mathrm{tot}}(t)$ is computed over the relevant time window, requiring $\mathcal{O}(N_t)$ time steps to resolve the system dynamics. To construct the full spectrum, this procedure is repeated for $N_\omega$ sampled frequency points. As a result, the total computational cost scales as $\mathcal{O}(N_t N_\omega)$. This approach circumvents the explicit evaluation of two-time correlation functions and avoids the double time integration over $(t',\tau)$ required in Eq.~(\ref{eq:SoPhysicalSpectrum}).

\begin{figure}[htbp]
    \centering
    \includegraphics[width=0.5\textwidth]{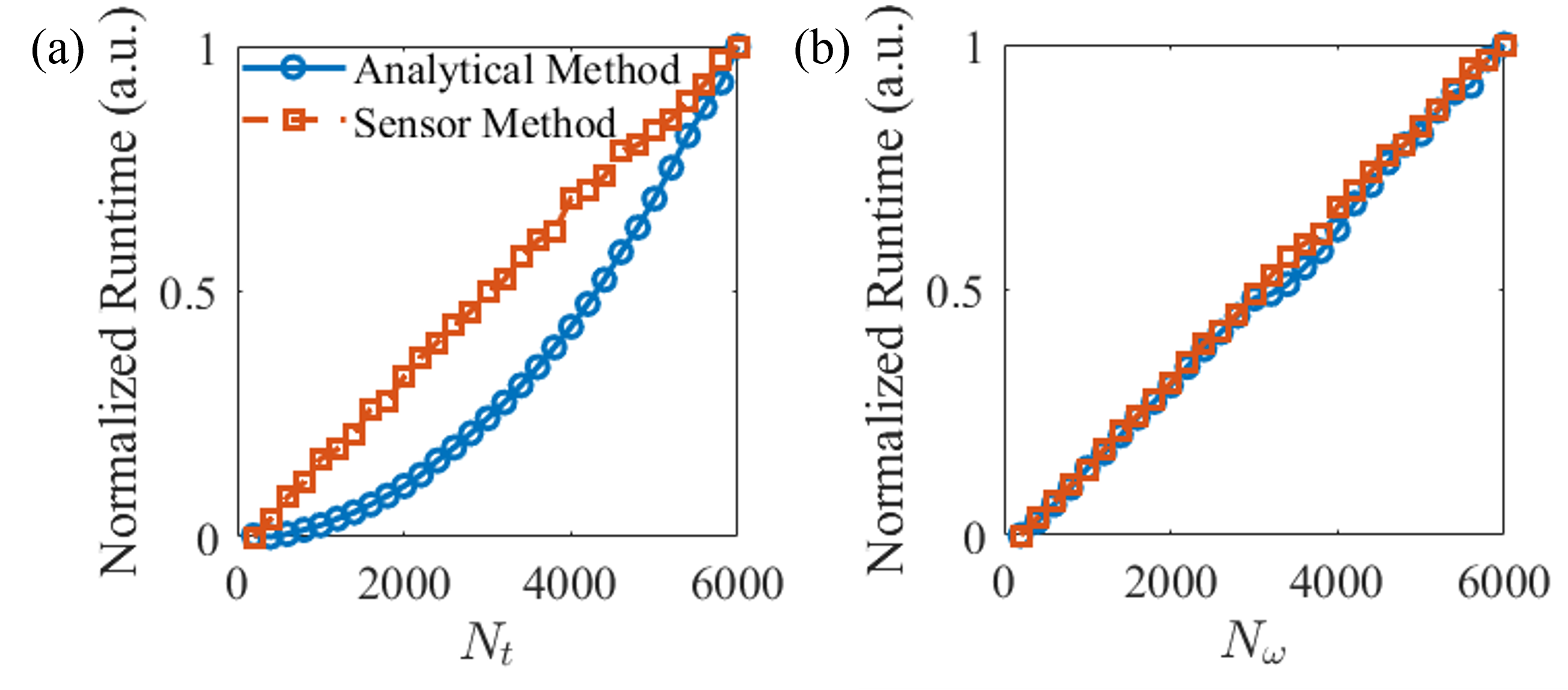}
\caption{Numerical runtime benchmark of the analytical method and the sensor method. 
The runtime is normalized to the range $[0,1]$ for visualization purposes only. 
(a) Runtime as a function of $N_t$ with $N_\tau=N_t$. 
(b) Runtime as a function of $N_\omega$.
}
    \label{fig:runtime_comparison}
\end{figure}

To further illustrate the computational scaling, we perform a numerical runtime benchmark for both methods, as shown in Fig.~\ref{fig:runtime_comparison}. We choose $N_\tau = N_t$, corresponding to the commonly used case where the delay-time resolution follows the physical-time discretization. The results show that the runtime of the analytical method exhibits a quadratic dependence on $N_t$, consistent with the scaling $\mathcal{O}(N_t^2 N_\omega)$ in this setting. In contrast, the sensor method displays a linear dependence on both $N_t$ and $N_\omega$, consistent with a scaling of $\mathcal{O}(N_t N_\omega)$. These numerical results are in good agreement with the scaling analysis presented above.

\begin{figure}[t]
\centering
\includegraphics[width=0.5\textwidth]{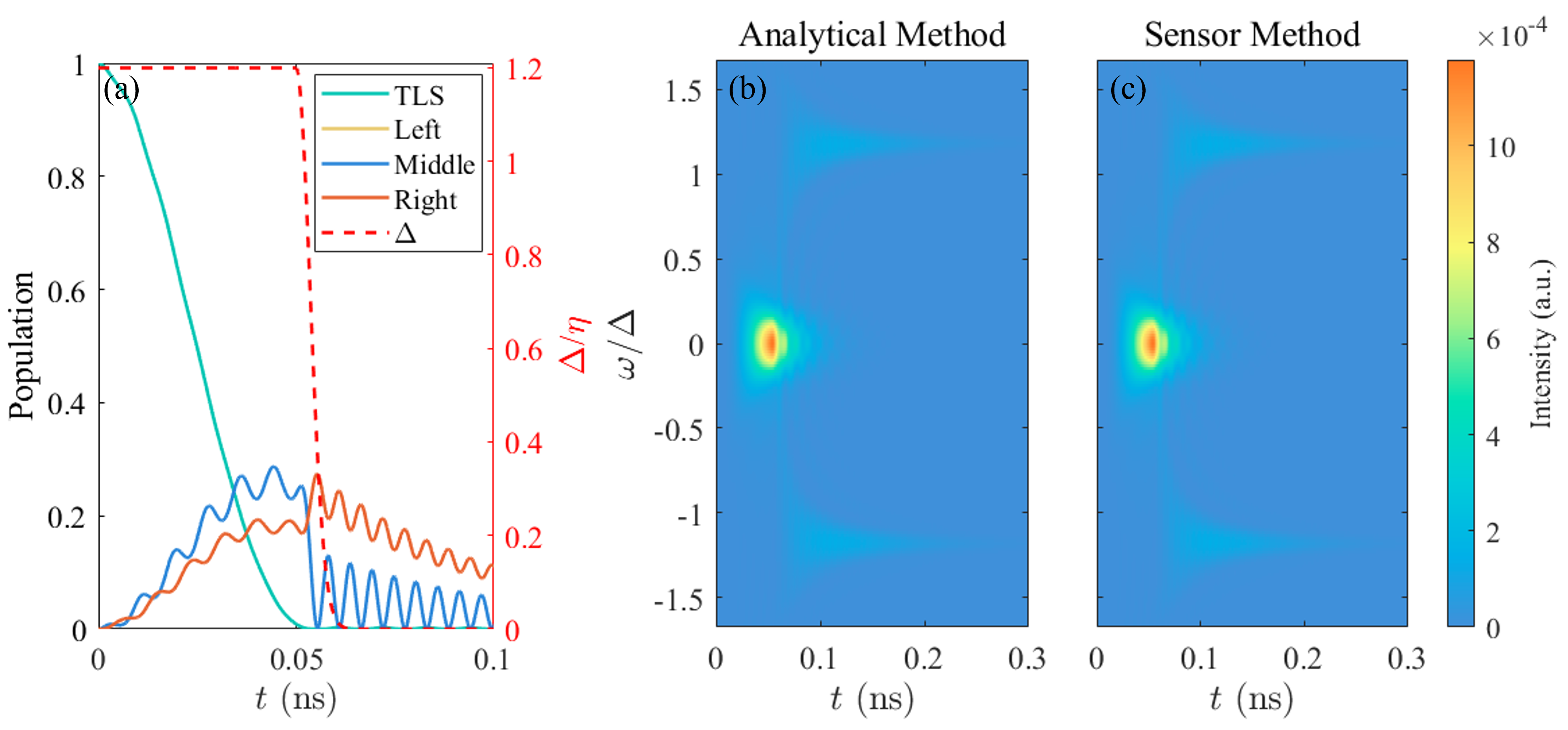}
\caption{
Validation of the sensor method under continuously time-dependent modulation.
(a) Smooth switching of the detuning with a finite fall time of 3 ps (dashed line), together with the population dynamics of the TLS and cavity modes.
(b)-(c) Comparison of the TRPS obtained from the analytical definition in Eq.~(\ref{eq:SoPhysicalSpectrum}) and the sensor method.
}
\label{fig:smooth_switch}
\end{figure}

To validate the applicability of the sensor method to general time-dependent systems, we consider a smooth modulation of the detuning with a Gaussian fall profile of width 3 ps, as shown in Fig.~\ref{fig:smooth_switch}(a). The parameters are chosen as $\delta_s = 50~\mathrm{GHz}$ and $\epsilon_k = 0.001~\mathrm{GHz}$, while all other system parameters are identical to those used in Fig.~\ref{fig:rabi_abcd}(b).

The corresponding TRPS are presented in Fig.~\ref{fig:smooth_switch}(b)-(c). Here, the analytical result is obtained by directly evaluating the definition of the TRPS in Eq.~(\ref{eq:SoPhysicalSpectrum}), which involves the calculation of two-time correlation functions and the subsequent double time integration. The analytical and sensor method remain in excellent agreement over the entire time and frequency range. 

\subsection{Multi-modal spectral dynamics}

\begin{figure*}[!t]
	\includegraphics[width=1.0\textwidth]{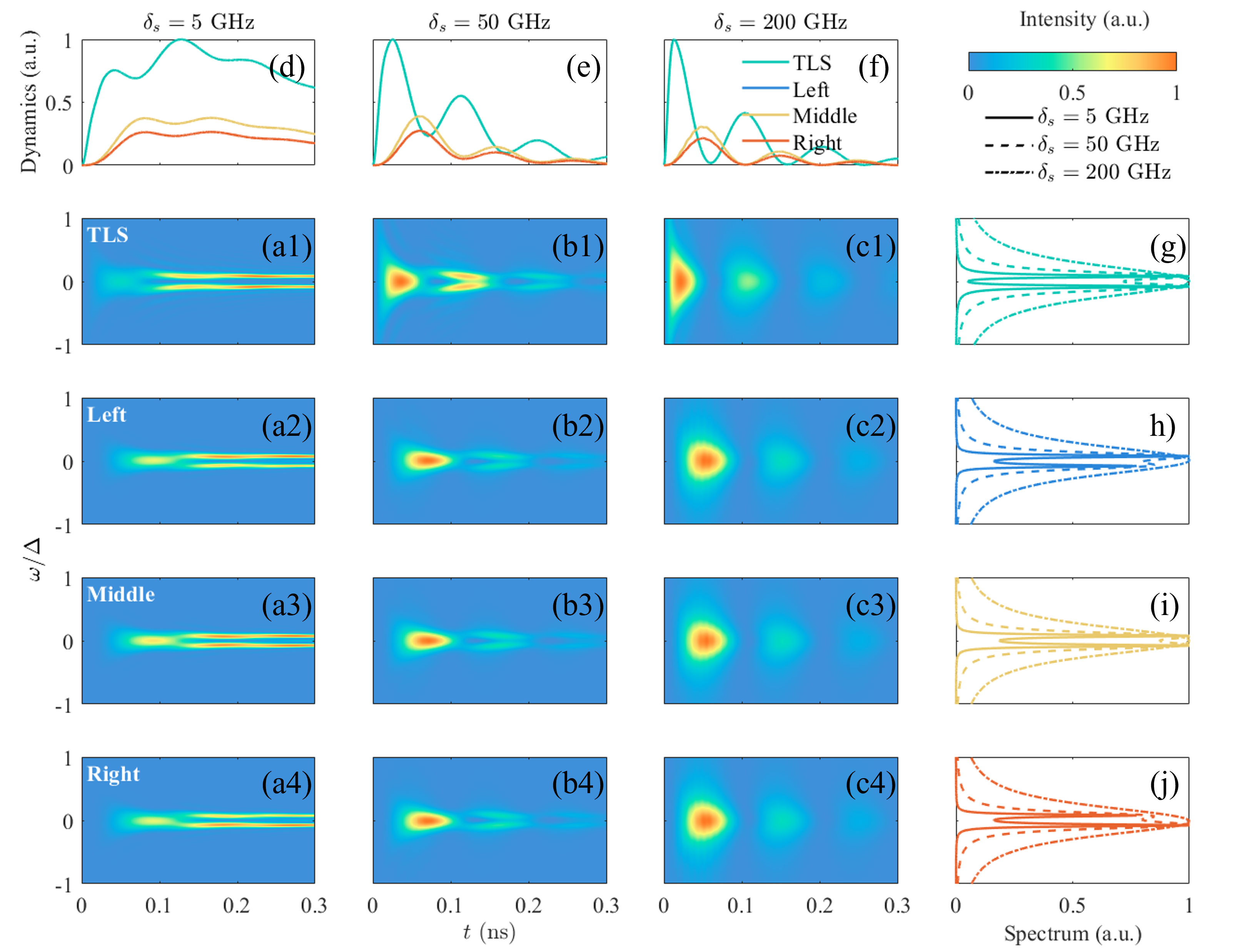}
	\caption{Time-resolved physical spectrum of the TLS-three-cavity system. Diagram of different resolution parameters, where $\delta_s=5~\text{GHz}, 50~\text{GHz}, 200~\text{GHz}$ are considered in (a), (b) and (c), respectively. Figs. (d), (e), (f) exhibit the frequency-integrated intensities versus time and (g), (h), (i), (j) show the time-integrated intensities versus frequency. In (a) and (b), after the first peak, the Rabi doublets emerge obviously. In (c) the doublets disappear while the frequency-integrated intensities are closed to the Rabi oscillation.}
	\label{fig:spec_1}
\end{figure*}

For the sake of consistency between experiments and theory, TRPS of the multi-modal dynamics in several cases and resolutions have been calculated, where the uncertainty relation is better demonstrated. The parameters for calculation are chosen to be the same as those used in the last section under the resonant condition. Fig.~\ref{fig:spec_1} shows the TRPS of the system in the no-switching case, calculated from the full system dynamics governed by Eq.~\eqref{eq:matrixH} together with the corresponding dissipative terms. Three spectral resolutions are considered, $\delta_s = 5~\text{GHz}, 50~\text{GHz}, 200~\text{GHz}$, while all other parameters are identical to those in Fig.~\ref{fig:rabi_abcd}(a). The sensor-system coupling strength is set to $\epsilon_k = 0.001~\mathrm{GHz}$, which is much smaller than all relevant energy scales of the system, ensuring that the sensors operate in the weak-coupling regime and do not perturb the intrinsic system dynamics. Panels (a1-c4) display the TRPS for the TLS and the three cavities, while panels (g-j) present the corresponding time-integrated spectrum, representing what would be observed by a spectrometer with the specified resolution. With high spectral resolution ($\delta_s = 5$~GHz), the Rabi doublet with splitting 
$\sqrt{2}g_{\mathrm{eff}} \approx 0.9 g$ is clearly visible, whereas for lower resolutions 
($\delta_s = 50$ and $200$~GHz) the doublet becomes increasingly blurred and eventually disappears, as the finite filter width washes out the frequency components. 

The TRPS reveals the Rabi dynamics in a manner strongly dependent on $\delta_s$, providing a direct visualization of the time-energy uncertainty relation. High spectral resolution yields sharp spectral peaks but suppresses oscillatory features in the time domain, while coarse resolution preserves the Rabi oscillation but merges the spectral doublet into a single broadened peak. In other words, improved spectral resolution necessarily degrades temporal resolution, and vice versa, as dictated by the time-energy uncertainty relation.

An additional feature arises from causality, the Rabi doublet in Fig.~\ref{fig:spec_1}(a) appears only after the field correlations have completed at least one full oscillation within the detection window. Before such temporal information becomes available, the spectrum necessarily exhibits a single peak because the spectrometer cannot infer the presence of oscillations prior to actually detecting them. This causal effect has no analogue in steady-state spectrum~\cite{yamaguchi2022time}. Consequently, even though the steady-state Rabi spectrum contains no component at $\omega=\omega_0$, the TRPS predicts that a high-resolution spectral filter can still detect an early-time pulse at the central frequency, which is a feature that would be completely missing in steady-state analysis due to the absence of measurement considerations.

\begin{figure*}[!t]
	\includegraphics[width=1.0\textwidth]{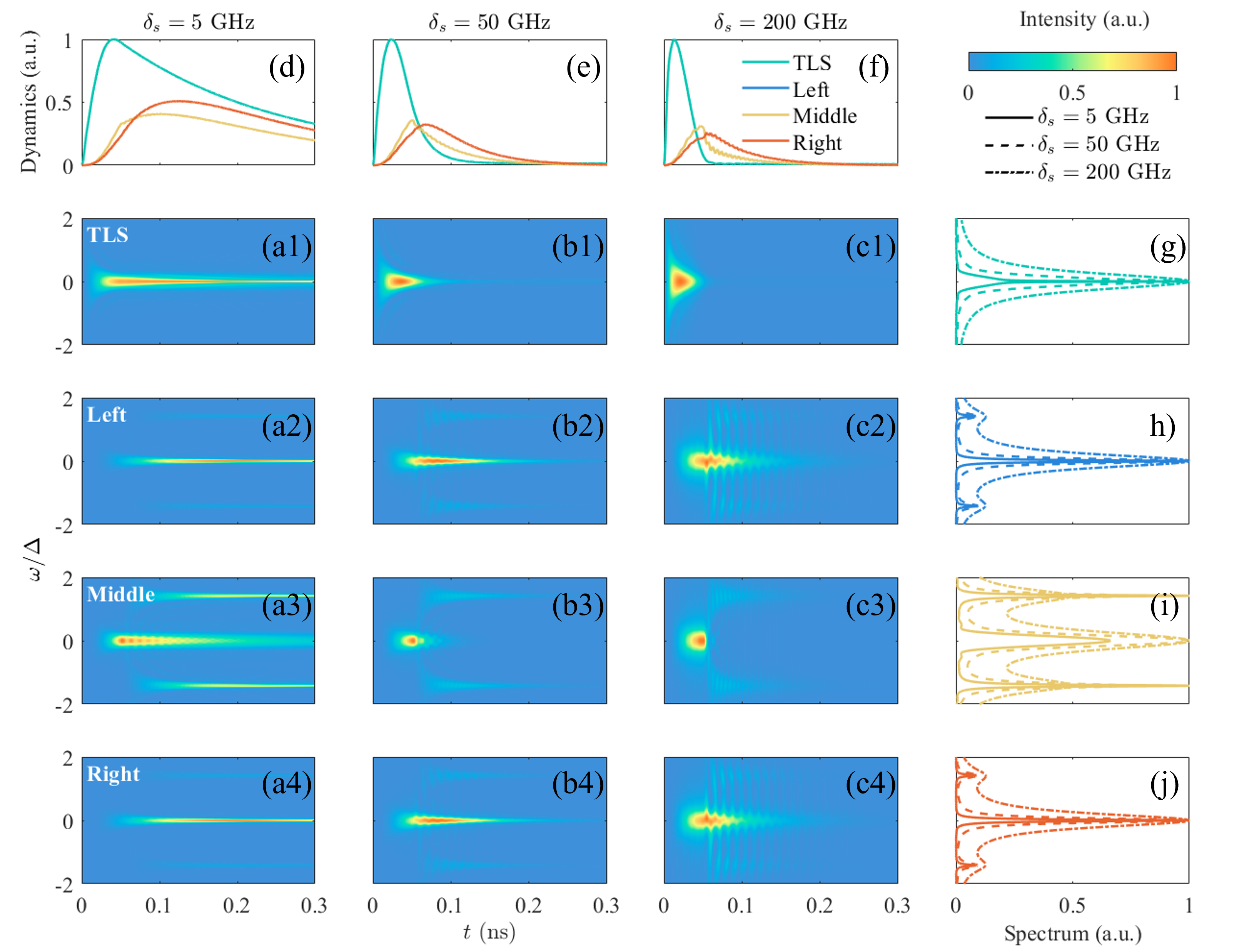}
	\caption{(a), (b) and (c) exhibit the TRPS for the case of switching off at the first valley of TLS in three different $\delta_s$, respectively. Figs.~(d), (e), (f) show the frequency-integrated intensities versus time and (g), (h), (i), (j) present the time-integrated intensities versus frequency. Since the oscillation is switched off before the first Rabi peak, the Rabi doublets are not resolved in the spectrum. Instead, the upper and lower supermodes emerge after the switch-off, appearing as two spectral lines in panel (a2)-(a4). As the filter bandwidth increases, interference beating between these supermodes becomes visible, as shown in panels (b) and (c).}
	\label{fig:spec_2}
\end{figure*}

\begin{figure*}[!t]
	\includegraphics[width=1.0\textwidth]{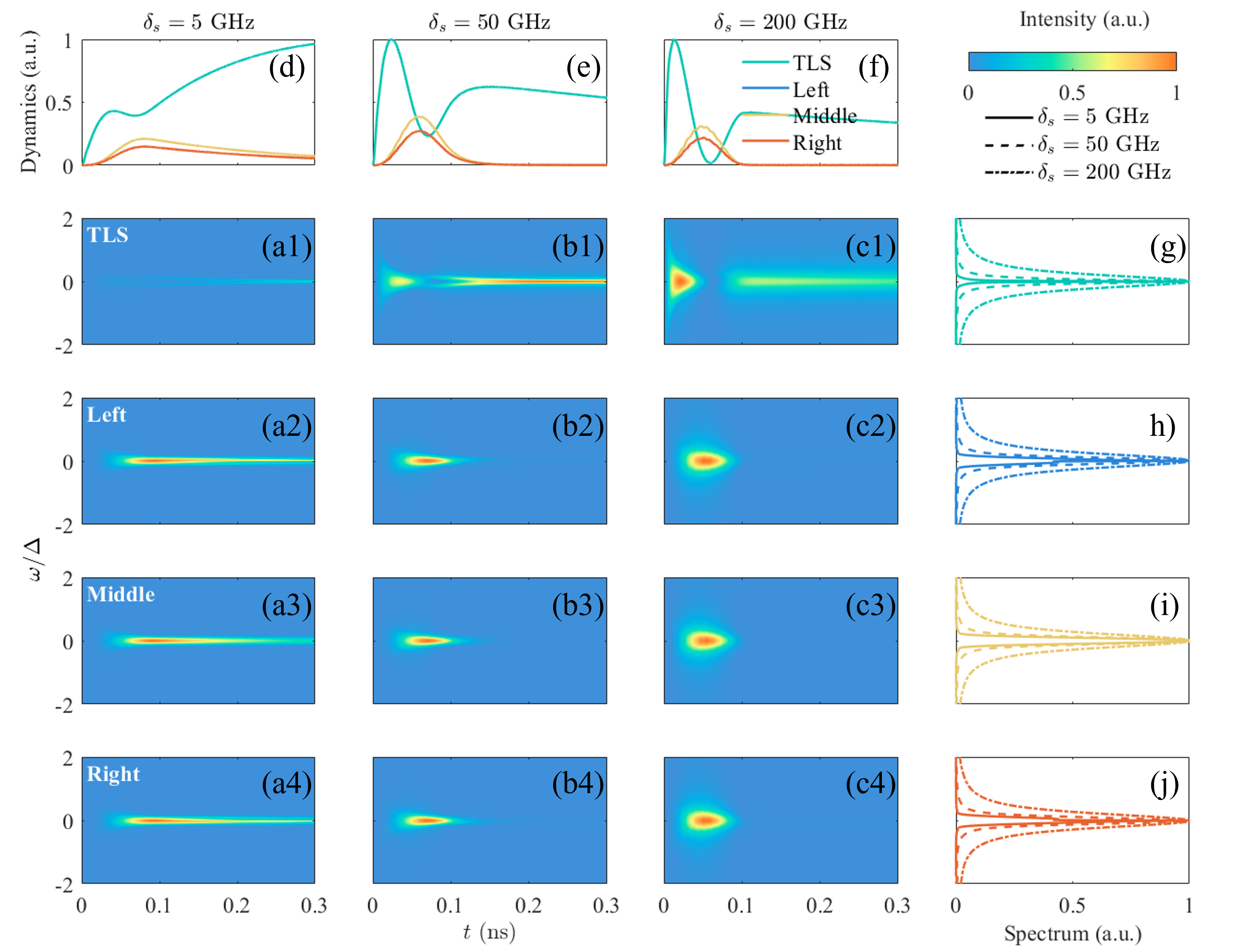}
	\caption{(a), (b) and (c) show the TRPS for the case of switching off at the first peak. Figs. (d), (e), (f) exhibit the frequency-integrated intensities versus time and (g), (h), (i), (j) present the time-integrated intensities versus frequency. Since the system is switched off after the first Rabi peak, the doublets briefly emerge in panels (a1)-(b1). }
	\label{fig:spec_3}
\end{figure*}

Fig.~\ref{fig:spec_2} and \ref{fig:spec_3} present the TRPS in the two switching-off scenarios, where the detuning is switched to $\Delta=0$ at the first valley (Fig.~\ref{fig:rabi_abcd}(b)) and at the first peak (Fig.~\ref{fig:rabi_abcd}(c)) of the TLS population, respectively. These two cases illustrate distinct transient behaviors arising from different distributions of excitation in the system at the switching moment.

% ======================  (1) Valley-off  ===========================
\paragraph*{Switching off at the first valley.}
In Fig.~\ref{fig:spec_2}(c1), the TRPS of the TLS exhibits a truncated response, reflecting the termination of the Rabi oscillation at the moment when the TLS is nearly unoccupied.
As the spectral resolution $\delta_s$ is reduced, the TRPS becomes extended along the time axis and concentrated along the frequency axis, consistent with the expected energy-time uncertainty. The TRPS of the three-cavity modes, shown in Fig.~\ref{fig:spec_2}(a2-a4), develops three distinct spectral lines immediately after the switching. These correspond to the eigenfrequencies of the TLS-three-cavity system at $\Delta=0$. Before switching, the zero-energy mode is primarily located in the middle cavity, but once $\Delta$ is set to zero, the zero-energy mode shifts entirely to the two lateral cavities. This redistribution is clearly visible in Fig.~\ref{fig:spec_2}(b2)-(b4): the central spectral line (zero-energy mode) disappears from the middle cavity after switching, while it appears in the two lateral cavities. The upper and lower supermode lines remain weak but present in all three cavities. For large $\delta_s$, the frequency axis becomes smooth, but the time-domain traces show rapid oscillations in the lateral cavities, as shown in Fig.~\ref{fig:spec_2}(b2)-(b3), originating from interference between the three supermode components.  

% ======================  (2) Peak-off  ===========================
\paragraph*{Switching off at the first peak.}
Fig.~\ref{fig:spec_3} shows the TRPS when the detuning is switched off at the first TLS peak. At this moment, the excitation resides almost entirely in the TLS. In this case only a single pulse of zero-energy mode emission appears in the cavities, as seen in Fig.~\ref{fig:spec_3}(c2)-(c4). This pulse corresponds to the zero-energy mode spectral line visible in Fig.~\ref{fig:spec_3}(a2)-(a4). The TLS TRPS behaves differently in this case. Because the system is switched off at a population maximum, the second half of the Rabi cycle is partially visible, and a short-lived Rabi doublet still appears temporarily in Fig.~\ref{fig:spec_3}(a1). Once the coupling to the cavities is removed, the TLS spectrum gradually narrows, as shown in Fig.~\ref{fig:spec_3}(c1), reflecting the absence of further energy exchange with the cavity modes.

As a short discussion on the implementation of ultrafast control, the TLS-three-cavity system can be realized in semiconductor quantum dots (QDs) and microcavities, where the coupling constant $g$, $\eta$ and the dissipation rate $\gamma$,$\kappa$ are well within the parameter range of current experiment techniques \cite{jin2014ultrafast, pellegrino2018deterministic}. The three-cavity scheme is experimentally more demanding than simpler configurations such as two-cavity \cite{jin2014ultrafast} or cavity-waveguide \cite{bradford2013spontaneous} systems, as it requires controlled intercavity coupling and coordinated tuning of multiple cavity modes. In contrast, these simpler systems typically control emission properties through modification of decay channels. The present scheme, however, enables selective control of the effective light-matter coupling via supermode redistribution without significantly altering the intrinsic decay processes. To support our simulation results, the speed of the sudden change in the detuning $\Delta$ needs to exceed the fastest quantity in the system, given by the coupling strength between cavities $\eta$ to perform ultrafast control. For instance, ultrafast control close to Terahertz range has been observed by utilizing free-carrier induced effects \cite{peinke2021tailoring,sattler2020probing}, which is suitable for the switching. The resulted transient Rabi doublets and their temporal broadening can be characterized using standard grating-based spectral filtering techniques \cite{peinke2021tailoring, sattler2020probing, jin2014ultrafast, fuhrmann2011dynamic}, making the proposed scheme directly compatible with existing ultrafast cQED measurement platforms. This capability opens the door to applications such as ultrafast quantum state control, on-demand modulation of vacuum Rabi splitting, and time-resolved engineering of emission spectrum.

\section{Conclusions}

In summary, we have developed and validated an efficient sensor-based numerical framework for computing time-resolved spectrum in nonstationary cQED systems. Using a three-cavity architecture as a minimal yet nontrivial testbed, we demonstrate that this multi-modal, time-domain methodology captures ultrafast, nonlocal modulation of light-matter coupling, where the transient detuning of the lateral cavities dynamically reshapes the local density of states experienced by the emitter without directly perturbing its transition frequency. Using the method, we compute the TRPS of the system with high numerical efficiency. The TRPS offers direct access to spectral signatures of mode conversion, following a switching event, the emission reorganizes into the instantaneous supermode basis, revealing the relocation and redistribution of the zero-energy mode.

By varying the spectral resolution, the TRPS further uncovers the operational manifestation of the time-energy uncertainty relation. High spectral resolution resolves transient Rabi doublets but suppresses temporal oscillations, whereas coarse resolution restores the Rabi dynamics at the expense of spectral detail. This provides a unified picture linking switching speed, mode formation, and the fundamental limits imposed by joint temporal-spectral characterization-an aspect of practical relevance because experimental control typically relies on frequency filtering.

Taken together, these results demonstrate that TRPS is a powerful diagnostic tool for the ultrafast control of multimode cQED systems, offering insights which are inaccessible by population dynamics alone. Our work establishes a formulation of the sensor method that is applicable to nonstationary dynamics, enabling efficient characterization of time-resolved spectral properties in quantum optical systems. This approach can be readily applied to more complex multimode architectures and dynamically reconfigurable photonic platforms.

\begin{acknowledgments}
This work is supported by the Zhejiang Province Leading Geese Plan (2024C01105),  the National Future Industry Innovation Mission, the Beijing Natural Science Foundation (L248103), the National Key Research and Development Program of China (2021YEB2800500), the National Natural Science Foundation of China (61574138, 61974131), and the Innovation Program for Quantum Science and Technology (2023ZD0300300).
\end{acknowledgments}

\appendix
\section{Sensor method}\label{app:sensormethodproof}
Del Valle \textit{et al.} have shown that the sensor method correctly reproduces the steady-state spectrum for time-independent systems~\cite{del2012theory}. Here we give a detailed explanation of the sensor method in the calculation of the TRPS. In contrast to Ref.~\cite{del2012theory}, we relax the assumption of time-independent dynamics and show that the method applies equally well to general nonstationary systems.

The physical spectrum of mode $O$ is written in Eq.~\ref{eq:SoPhysicalSpectrum} \cite{yamaguchi2022time, del2012theory}, where the expectation value $\langle O^\dagger(t')\,O(t'-\tau)\rangle$ is taken over the reduced system state after tracing out the external environment.

We start from a (possibly time-dependent) system Hamiltonian $H_S(t)$ and weakly couple a damped sensor to the system. The Hamiltonian of the sensor is $\omega_k \zeta_k^\dagger \zeta_k$, where the transition frequency $\omega_k$ is the detection frequency of interest and the sensor linewidth is given by $\Gamma$. The interaction Hamiltonian between the sensor and the system operator $O$ reads $H_{\mathrm{int}} = \epsilon_k (\zeta_k^\dagger O + O^\dagger \zeta_k)$, where the couplings $\epsilon_k$ are assumed to be sufficiently small so as not to perturb the intrinsic system dynamics. For clarity, we present the derivation for a single sensor at a fixed frequency $\omega_k$. The full spectrum is then obtained by scanning $\omega_k$ over the desired frequency range.

In the corresponding Heisenberg picture, we may additionally include a reservoir $R$ to $H_S(t)$ to model the system as an open quantum system.

The Heisenberg equation for $\zeta_k$ is 
\begin{equation}
\frac{d\zeta_k}{dt}
= i[H_{\mathrm{tot}}(t),\zeta_k] - \frac{\Gamma}{2}\zeta_k,
\label{eq:HeisenbergZeta}
\end{equation}
where $H_{\mathrm{tot}}(t)$ is the sensors-system joint Hamiltonian shown in Eq.~\eqref{eq:Htot}.

Since the sensor operator $\zeta_k$ acts on a different Hilbert space, it commutes with all bath operators in $R$ as well as with $H_S(t)$. Consequently, these terms drop out of the commutator $i[H_{\mathrm{tot}}(t),\zeta_k]$, yielding $i[H_{\mathrm{tot}}(t),\zeta_k]=i[H_{\mathrm{int}}+\omega_k\,\zeta_k^\dagger\zeta_k,\zeta_k]$. As a result, the sensor still obeys the simple damped-oscillator equation
\begin{equation}
\frac{d\zeta_k}{dt}
= -\Big(i\omega_k +\frac{\Gamma}{2}\Big)\zeta_k
  - i\epsilon_k\, O ,
\label{eq:Zeta}
\end{equation}
irrespective of whether the system Hamiltonian is time dependent. The equations of motion for the first-order moments and the sensor populations then read
\begin{eqnarray}
  \frac{d\mean{\zeta_k}}{dt} &=& -\Big(i\omega_k +\frac{\Gamma}{2}\Big)\mean{\zeta_k}
  - i\epsilon_k \mean{O}, \\
  \frac{d\mean{\hat{n}_k}}{dt} &=& -\Gamma\,\mean{\hat{n}_k} +
  2\,\Re\!\left[i\epsilon_k\mean{O^\dagger \zeta_k}\right], \label{eq:NZeta}
\end{eqnarray}
where $\hat n_k=\zeta_k^\dagger \zeta_k$.

For a fixed detection time $t$ and delay $\tau\ge 0$, we first differentiate the sensor operator with respect to the delay,
\begin{equation}
\begin{aligned}
\frac{d}{d\tau}\zeta_k(t-\tau)
&= -\left. \frac{d}{dx}\zeta_k(x)\right|_{x \to t-\tau} \\
&= \Big(i\omega_k +\frac{\Gamma}{2}\Big)\zeta_k(t-\tau)
  + i\epsilon_k\,O(t-\tau),
\end{aligned}
\end{equation}
where we have used the chain rule together with Eq.~(\ref{eq:Zeta}). Multiplying on the left by $O^\dagger(t)$ and taking the expectation value, we obtain
\begin{equation}\label{eq:2TCorrelation}
\begin{aligned}
  \frac{d}{d\tau}\mean{O^\dagger(t) \zeta_k(t-\tau)}
  =& \Big(i\omega_k +\frac{\Gamma}{2}\Big)\mean{O^\dagger(t)\zeta_k(t-\tau)} \\
   & + i\epsilon_k \mean{O^\dagger(t)\,O(t-\tau)} .
\end{aligned}
\end{equation}
Equation~(\ref{eq:2TCorrelation}) therefore follows purely from the local Heisenberg equation of motion~(\ref{eq:Zeta}) for the sensor operator $\zeta_k$ and from the definition of the two-time correlator. In particular, it neither relies on the quantum regression theorem nor requires the dynamics generated by $H_S(t)$ to be Markovian, and any non-Markovian memory effects are fully contained in the system correlator $\langle O^\dagger(t)\,O(t-\tau)\rangle$ itself.

For notational convenience we introduce
\begin{equation}
\begin{aligned}
\mean{O^\dagger \zeta_k}_t(\tau)&\equiv\mean{O^\dagger(t) \zeta_k(t-\tau)}, \\
\mean{O^\dagger O}_t(\tau)&\equiv\mean{O^\dagger(t) O(t-\tau)} .
\end{aligned}
\end{equation}
Applying the Laplace transform with respect to $\tau$ to Eq.~(\ref{eq:2TCorrelation}) yields
\begin{equation}\label{DLap2Tcorr}
\begin{aligned}
  &s\,\mathscr{L}\{\mean{O^\dagger \zeta_k}_t(\tau)\}(s) - \mean{O^\dagger \zeta_k}_t(0) \\
  =&\,
  i\epsilon_k\,\mathscr{L}\{\mean{O^\dagger O}_t(\tau)\}(s)
  +
  \Big(i\omega_k+\frac{\Gamma}{2}\Big)\mathscr{L}\{\mean{O^\dagger \zeta_k}_t(\tau)\}(s) .
\end{aligned}
\end{equation}
Evaluating this relation at $s_0=i\omega_k+\Gamma/2$ we obtain
\begin{equation}\label{DLap2TcorrSimplified}
  \mean{O^\dagger \zeta_k}_t(0)
  = -i\epsilon_k\,\mathscr{L}\{\mean{O^\dagger O}_t(\tau)\}(s_0) .
\end{equation}
Using the integral representation of the Laplace transform, the mixed correlator at equal times can thus be written as
\begin{equation}\label{eq:Lap2Tcorr}
\begin{aligned}
  \mean{O^\dagger(t)\zeta_k(t)}
  &= -i\epsilon_k\,\mathscr{L}\{\mean{O^\dagger O}_t(\tau)\}(s_0) \\[2pt]
  &= -i\epsilon_k\int_{0}^{\infty}
  d\tau\,e^{-(i\omega_k+\frac{\Gamma}{2})\tau}
   \mean{O^\dagger O}_t(\tau).
\end{aligned}
\end{equation}
Solving Eq.~(\ref{eq:NZeta}) with the initial time taken at $t_0\to-\infty$, we obtain
\begin{equation}\label{eq:NZetaSolution}
\begin{aligned}
  \mean{\hat{n}_k(t)} 
= &\, e^{-\Gamma t}\,\mean{\hat{n}_k(-\infty)} \\ 
& + \int_{-\infty}^{t} dt'\, e^{-\Gamma (t-t')}\,
2\,\Re\!\left[i\,\epsilon_k\,\mean{O^\dagger(t')\,\zeta_k(t')}\right].
\end{aligned}
\end{equation}
Substituting Eq.~(\ref{eq:Lap2Tcorr}) into Eq.~(\ref{eq:NZetaSolution}) and comparing with the definition Eq.~(\ref{eq:SoPhysicalSpectrum}), we find that the population of the sensor mode consists of a contribution proportional to the physical spectrum of the mode $O$ plus a simple exponential transient,
\begin{equation}\label{eq:FullSensorDynamics}
  \mean{n_k(t)} = \frac{2\epsilon_k^2\pi}{\Gamma}\,S_O(\omega_k, t, \Gamma)
  + C\,e^{-\Gamma t},
\end{equation}
where $C$ is a constant determined by the initial value $\mean{n_k(-\infty)}$. In the absence of initial sensor excitations ($C=0$), the sensor dynamics directly reproduce the physical spectrum,
\begin{equation}\label{eq:SensorPhySpectra}
  \mean{n_k(t)} = \frac{2\epsilon_k^2\pi}{\Gamma}\,S_O(\omega_k, t, \Gamma).
\end{equation}
which is consistent with Eq.~\eqref{eq:nkSensorPhySpectra}. In the steady-state limit $t\rightarrow\infty$, the result reduces to the frequency-filtered power spectrum obtained in Ref.~\cite{del2012theory}.
% It is interesting to note that the conclusion holds even for a large $\epsilon_k$, depsite the fact that the auxiliary mode is involved in the dynamics of the system. This gives the insight that if the auxiliary mode (might be a quantumized mode of a local optical field) is the leaky channel of the system, the dynamics of the system would be reshaped by the linewidth and the frequency (especially in a large bath which the behavior of the auxiliary mode will not impact the system too much).

\section{Deriavtion of the supermodes}
\label{app:supermodes}
The full Hamiltonian of the system is Eq.~\eqref{eq:matrixH}. The three eigenvalues of the $3\times 3$ matrix in the lower right corner are Eq.~\eqref{eq:3cavEvec}. The $l_{-}$ and $l_{+}$ are normalizing factors of the two supermodes, which could be explicitly written as 
\bea
l_{\pm}=\sqrt{2}\sqrt{ \frac{    (  \Delta^2+2\eta^2  )   ( \Delta^2+\eta^2 \pm \Delta \sqrt{\Delta^2+2\eta^2}  )                    
	}{\eta^4}      
}
\eea
The full Hamiltonian after diagonalizing is written as (here the $v_i^{(j)}$ represents the $j$'s component of the  eigenvector $v_i$)

\bea
&&H_{dia} = \hat P ^{\dagger} \hat H \hat P \nn
&&=\left[\begin{array}{cccc}
	\omega_0&g v_{-}^{(2)} &  g v_{0}^{(2)}&g v_{+}^{(2)}\nn
	gv_{-}^{(2)}& \omega_0-\sqrt{\Delta^2+2\eta^2} &0&0\nn
	gv_{0}^{(2)}&0&\omega_0&0\nn
	gv_{+}^{(2)}&0&0&\omega_0+\sqrt{\Delta^2+2\eta^2}
\end{array}\right]~~\label{diaHHH}
\eea
with the transformation matrix $\hat P$
\bea
\hat P =\left[\begin{array}{cccc}
	1&0&0&0\nn
	0&v_{-}^{(1)}&v_{0}^{(1)}&v_{+}^{(1)}\nn
	0&v_{-}^{(2)}&v_{0}^{(2)}&v_{+}^{(2)}\nn
	0&v_{-}^{(3)}&v_{0}^{(3)}&v_{+}^{(3)}
\end{array}\right]
\eea

In the diagonalized full Hamiltonian one can see that the effective coupling constant between the TLS and the zero-energy mode is $g v_{0}^{(2)}$, which entirely depends on the second component of the eigenmode $\vec v_{0}$, and it is written explicitly as
\bea
g_{\mathrm{eff}} = \frac{\frac{\Delta}{\eta}}{\sqrt{2+(\frac{\Delta}{\eta})^2}} g 
\eea
By changing the detuning $\Delta$, the effective coupling constant between zero-energy mode and the TLS could be adjusted from $0$ to $g$. Compared to the two-cavity case, the range of $g_{\mathrm{eff}}$ is wider.

\section{Effective control regime of the three-cavity model set by $g/\eta$}
\label{app:g_eta}

As discussed previously, at $\Delta=0$ the central supermode is resonant with the TLS but has vanishing effective coupling, while the upper and lower supermodes remain detuned and only couple to the TLS through reduced amplitudes. In this situation, the effectiveness of suppressing Rabi oscillations depends on the ratio $g/\eta$, which controls the residual coupling to the detuned supermodes. 

\begin{figure}[!htbp]
    \centering
    \includegraphics[width=0.5\textwidth]{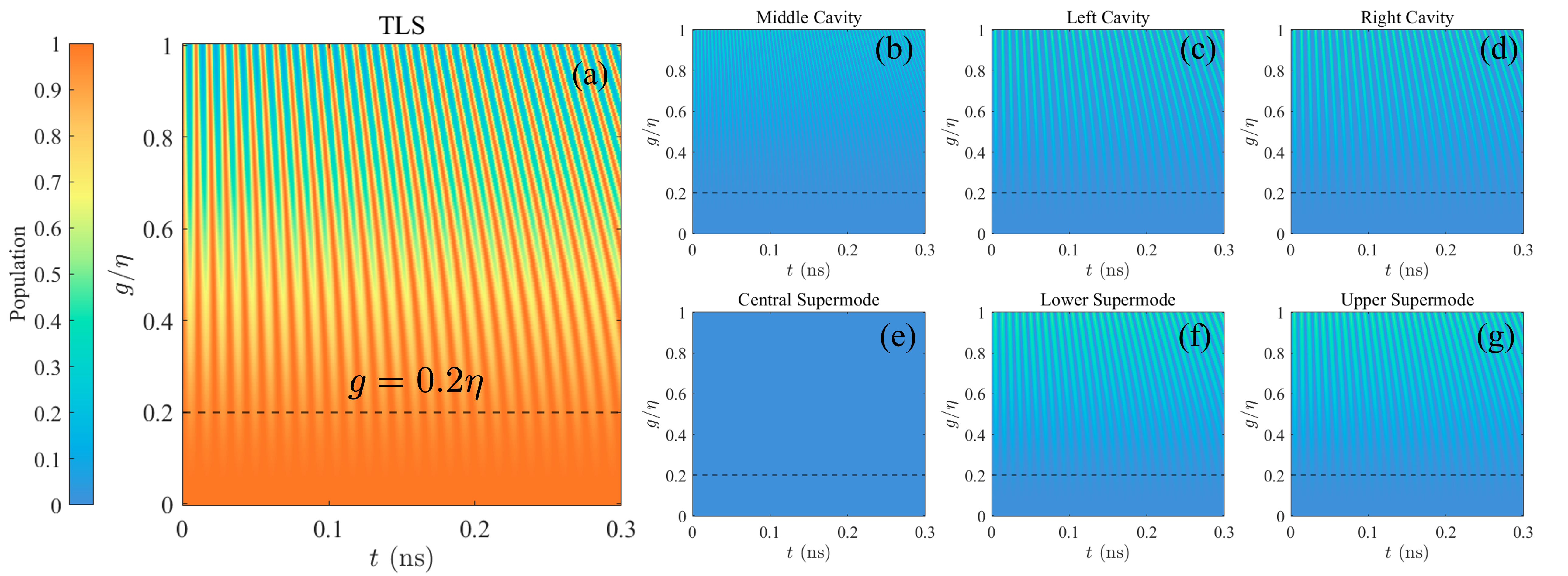}
    \caption{Population maps obtained by scanning $g$ at $\Delta=0$. The parameters are the same as in Fig.~\ref{fig:rabi_abcd}(a), except that dissipation is neglected.}
    \label{fig:g_scan_population}
\end{figure}

To directly visualize the impact on the system dynamics, we perform numerical simulations by scanning $g$ at $\Delta=0$, using the same parameters as in Fig.~\ref{fig:rabi_abcd}(a), except that dissipation is neglected. As shown in Fig.~\ref{fig:g_scan_population}, no pronounced Rabi oscillation is observed for $g \lesssim 0.2\eta$, whereas clear oscillatory dynamics emerges when $g$ becomes comparable to $\eta$.

\begin{figure}[!htbp]
\centering
\includegraphics[width=\linewidth]{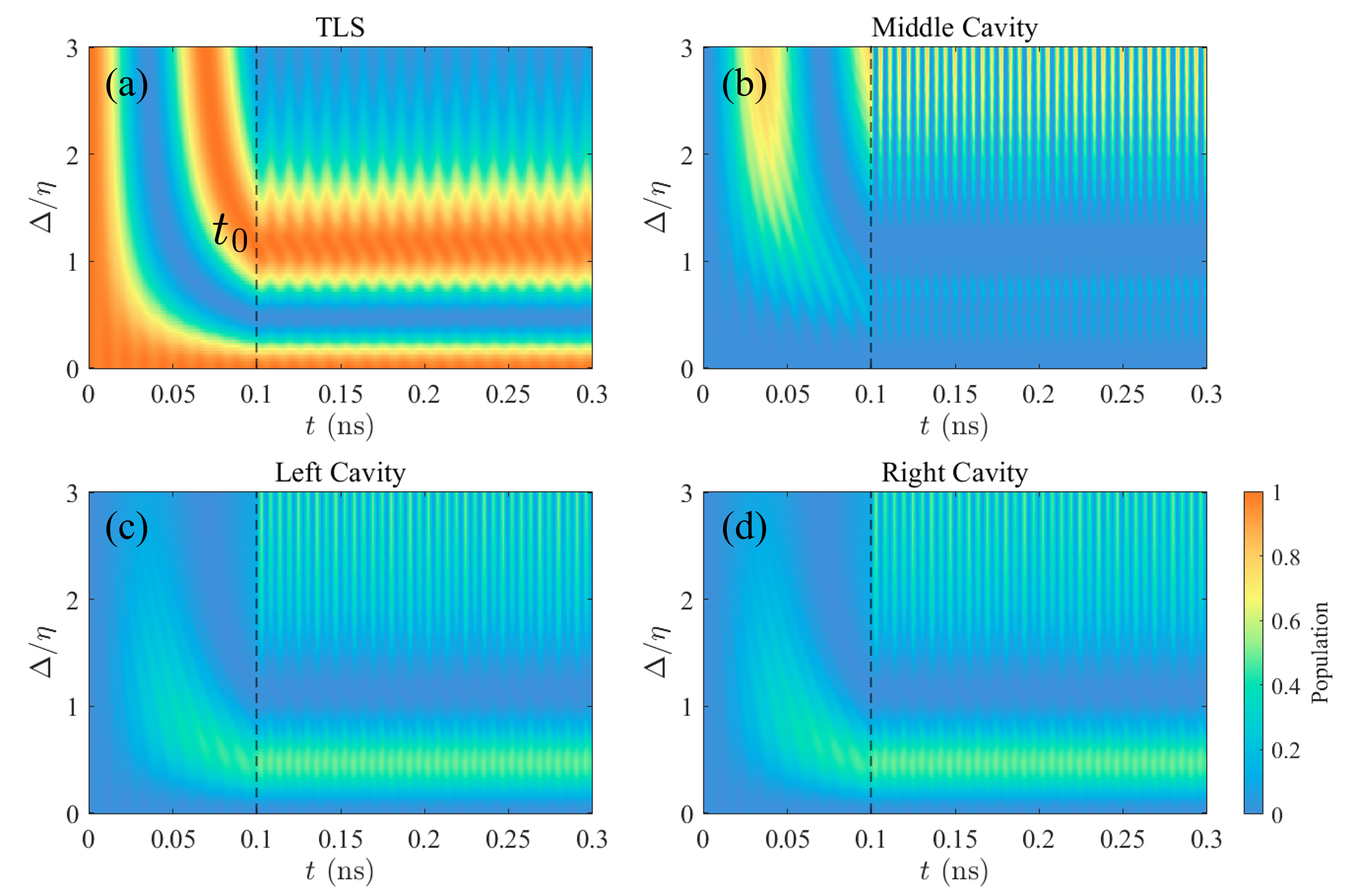}
\caption{
Population dynamics for different detuning amplitudes $\Delta$. 
Larger $|\Delta|$ leads to faster oscillations before the switching, while the dominant TLS Rabi oscillation remains suppressed after $\Delta \rightarrow 0$.
}
\label{fig:Delta_vs_population}
\end{figure}

\begin{figure}[!htbp]
\centering
\includegraphics[width=\linewidth]{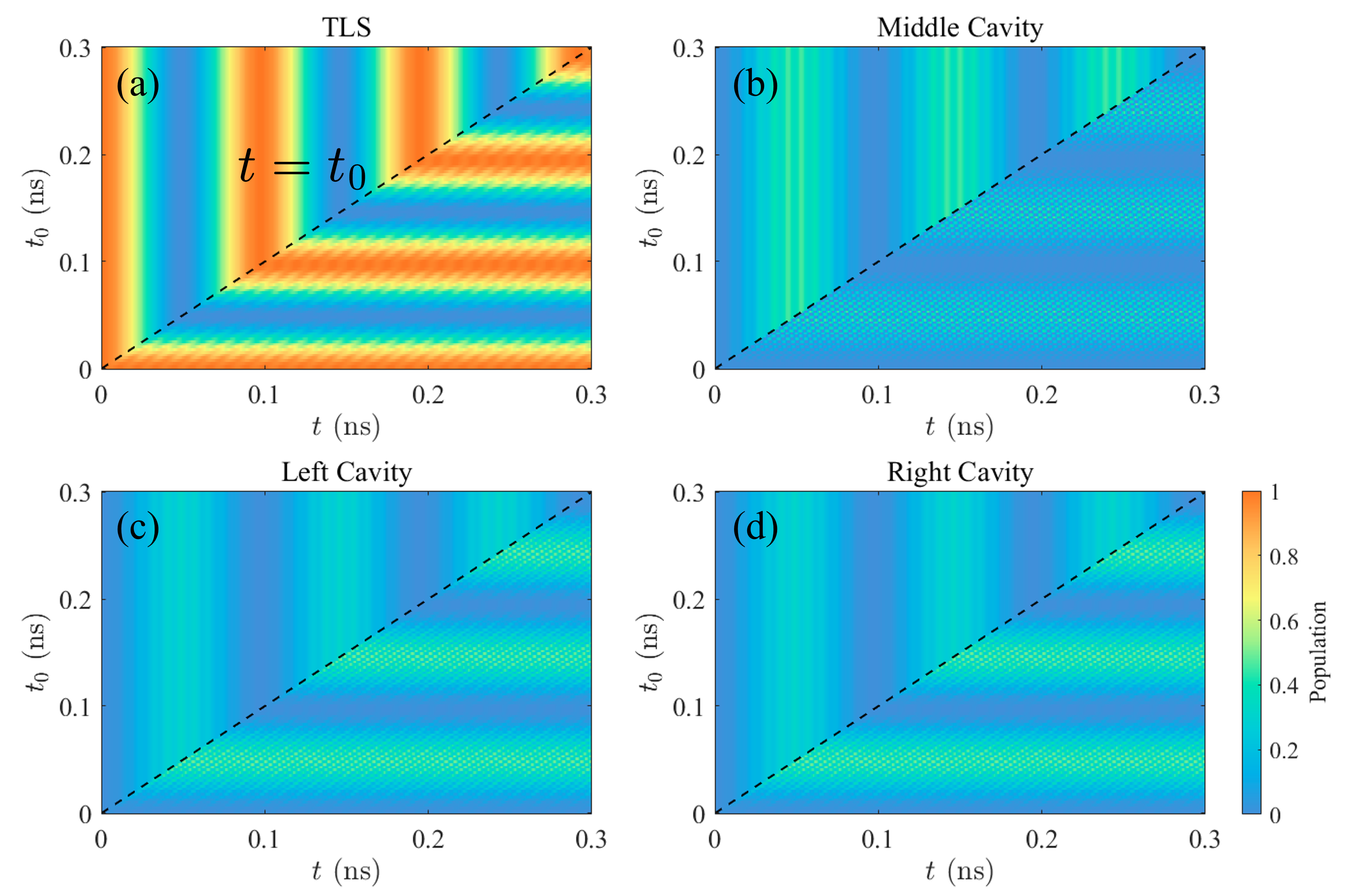}
\caption{
Population dynamics for different switching times $t_0$. 
The switching phase modifies the transient evolution, while the suppression of the dominant TLS Rabi oscillation persists after the switching.
}
\label{fig:t0_vs_population}
\end{figure}

\section{Robustness of control in the three-cavity model}
\label{app:delta_phase}

We examine the dependence of the switching dynamics on the detuning amplitude $\Delta$ and the switching time $t_0$ in the three-cavity model. The parameters are the same as those used in Fig.~\ref{fig:rabi_abcd}, unless otherwise specified.

As shown in Fig.~\ref{fig:Delta_vs_population}, increasing $|\Delta|$ leads to faster Rabi oscillations before the switching, while after $\Delta \rightarrow 0$ the dominant TLS Rabi oscillation remains suppressed, with only weak residual dynamics arising from the off-resonant coupling to the upper and lower supermodes. 
Fig.~\ref{fig:t0_vs_population} shows that varying $t_0$ modifies the instantaneous state at the switching moment and thus the transient evolution, but the suppression of the dominant Rabi oscillation persists for different switching phases.

These results confirm that the control mechanism is robust against variations in detuning amplitude and switching phase, as it relies on the vanishing coupling to the central supermode and the large detuning of the upper and lower supermodes at $\Delta=0$.

\bibliography{reference}

\end{document}